\documentclass[sigconf]{acmart}
\AtBeginDocument{%
  }

\copyrightyear{2025}
\acmYear{2025}
\setcopyright{rightsretained}
\acmConference[MM '25]{Proceedings of the 33rd ACM International Conference on Multimedia}{October 27--31, 2025}{Dublin, Ireland}
\acmBooktitle{Proceedings of the 33rd ACM International Conference on Multimedia (MM '25), October 27--31, 2025, Dublin, Ireland}
\acmDOI{10.1145/3746027.3755619}
\acmISBN{979-8-4007-2035-2/2025/10}
\begin{document}

\title{HER2 Expression Prediction with Flexible Multi-Modal Inputs via Dynamic Bidirectional Reconstruction}

% %%
% %% The "author" command and its associated commands are used to define
% %% the authors and their affiliations.
% %% Of note is the shared affiliation of the first two authors, and the
% %% "authornote" and "authornotemark" commands
% %% used to denote shared contribution to the research.
\author{Jie Qin}
\authornote{Both co-first authors contributed equally to this research.}
\affiliation{
  \institution{School of Computer Science and Technology, University of the Chinese Academy of Sciences}
  \city{Beijing}
  \country{China}
}
\email{2214524977@stu.xjtu.edu.cn}

\author{Wei Yang}
\authornotemark[1]  % 引用第一个authornote的标记
\affiliation{
  \institution{School of Computer Science and Technology, University of the Chinese Academy of Sciences}
  \city{Beijing}
  \country{China}
}
\email{220231091125@ncepu.edu.cn}

\author{Yan Su}
\affiliation{
  \institution{School of Computer Science and Technology, University of the Chinese Academy of Sciences}
  \city{Beijing}
  \country{China}
}
\email{suyan@mail.bnu.edu.cn}

\author{Yiran Zhu}
\affiliation{
  \institution{School of Computer Science and Technology, University of the Chinese Academy of Sciences}
  \city{Beijing}
  \country{China}
}
\email{220231091129@ncepu.edu.cn}

\author{Weizhen Li}
\affiliation{
  \institution{School of Computer Science and Technology, University of the Chinese Academy of Sciences}
  \city{Beijing}
  \country{China}
}
\email{19100750506@163.com}

\author{Yunyue Pan}
\affiliation{
  \institution{School of Computer Science and Technology, University of the Chinese Academy of Sciences}
  \city{Beijing}
  \country{China}
}
\email{2100016253@stu.pku.edu.cn}

\author{Chengchang Pan}
\authornote{Corresponding author.}
\affiliation{
  \institution{School of Computer Science and Technology, University of the Chinese Academy of Sciences}
  \city{Beijing}
  \country{China}
}
\email{166353314@qq.com}

\author{Honggang Qi}
\authornotemark[2]
\affiliation{%
  \institution{School of Computer Science and Technology, University of Chinese Academy of Sciences}
  \city{Beijing}
  \country{China}
}

\email{hgqi@ucas.ac.cn}


\begin{abstract}
In the field of HER2 expression level assessment for breast cancer, clinical evaluations often rely on the synergistic analysis of both H\&E and IHC stained images. However, acquiring dual-modality images for the same patient is frequently hindered by complex clinical workflows and high costs, resulting in missing modalities. To address this challenge, we propose an adaptive bimodal input prediction framework that flexibly supports both single-modality and dual-modality inputs. This framework employs a dynamic branch selection mechanism to overcome the rigid dependency of existing models on complete inputs, enabling accurate predictions using either H\&E or IHC images alone, while retaining the ability for joint inference when both modalities are available. The core technical innovations include: a missing modality branch selector that dynamically activates either a modality completion process or an end-to-end dual-modality inference pipeline based on the available input; and a cross-modal generative adversarial network (CM-GAN) that facilitates context-aware reconstruction of the missing modality in the feature space. This design improves the prediction accuracy from 71.44\% to 94.25\% when using single-modality H\&E images, significantly mitigating performance degradation caused by incomplete information. Experimental results demonstrate that the proposed framework achieves a prediction accuracy of 95.09\% with full dual-modality input and maintains a high reliability of 90.28\% under single-modality conditions. By adopting this “dual-modality preferred, single-modality compatible” flexible architecture, healthcare institutions can achieve near dual-modality accuracy without mandating synchronized acquisition of both image types. This is particularly valuable for regions with limited IHC staining infrastructure, offering a cost-effective clinical solution and substantially enhancing the accessibility of HER2 expression level assessment.

\end{abstract}




% %% A "teaser" image appears between the author and affiliation
% %% information and the body of the document, and typically spans the

\received{03 April 2025}
\received[revised]{12 April 2025}
\received[accepted]{06 July 2025}

\maketitle

\section{Introduction}
Breast cancer is the most common malignancy among women worldwide, and accurate evaluation of its molecular subtypes is critical for guiding personalized treatment strategies\cite{ref1}. Human epidermal growth factor receptor 2 (HER2) is a key biomarker in breast cancer, and its expression level directly influences the selection of targeted therapies\cite{ref2,ref3}. Therefore, precise assessment of HER2 status is of vital importance.

Currently, HER2 expression is primarily assessed through immunohistochemistry (IHC) and in situ hybridization (ISH). Although these techniques are widely used, they have several limitations. IHC scoring depends heavily on the subjective judgment of pathologists and is prone to significant inter-observer and inter-laboratory variability\cite{ref5}. The recently introduced "HER2-low" subtype further highlights the challenge of consistent interpretation for borderline cases\cite{ref4,ref13}. ISH techniques (including FISH and CISH) provide information on gene amplification but involve complex protocols and costly reagents, making them less accessible in resource-limited settings\cite{ref6}. Additionally, the standard IHC/ISH workflow often takes several days to deliver results, delaying clinical decision-making. These limitations have driven the development of artificial intelligence (AI)-based automated HER2 assessment approaches, aimed at enhancing objectivity and efficiency\cite{ref7,ref8}.

Early automation efforts focused on computer-assisted scoring of IHC images\cite{ref7}. In recent years, the application of deep learning techniques has further improved the consistency of HER2 evaluation\cite{ref8}. However, most existing AI models operate on a single imaging modality and thus inherit the inherent limitations of unimodal data. Hematoxylin and eosin (H\&E)-stained slides provide morphological context but lack protein expression information, while IHC directly visualizes HER2 protein distribution but can be affected by staining variability. Several studies have attempted to predict HER2 status using either H\&E or IHC images alone[10,11]; although these approaches achieved promising results, their overall accuracy and robustness remain constrained by incomplete modality information\cite{ref12}.

In principle, combining H\&E and IHC images can provide complementary morphological and molecular insights, potentially overcoming the limitations of single-modality approaches. Multimodal fusion has shown notable potential in other oncology tasks\cite{ref17}. For instance, McKinney et al. developed a breast cancer screening system that surpassed radiologist performance by integrating multi-view mammograms with clinical data\cite{ref18}, while Joo et al. achieved accurate prediction of pathological complete response to neoadjuvant chemotherapy by combining MRI and clinical information\cite{ref19}.

However, most existing multimodal AI frameworks assume the availability of all modalities, which is often not feasible in real-world clinical settings. For example, H\&E slides are readily available in routine diagnosis, but corresponding IHC slides may be missing due to limited tissue, time constraints, or cost. Current methods lack mechanisms to effectively handle missing modalities—if an input is incomplete, the model may fail entirely or rely on naive imputation (e.g., zero-filling or mean substitution), which can distort feature distributions and degrade performance. Moreover, existing multimodal fusion strategies typically use fixed weighting schemes that cannot adapt to variations in image quality. For instance, when an IHC image is overstained or faint, the model should rely more on H\&E data—but static fusion methods cannot make such adjustments flexibly.

To address these challenges, we propose a novel HER2 prediction framework that is both multimodal and modality-flexible. Our approach not only fully leverages the advantages of multimodal information but also effectively handles modality absence and quality variation. The core contributions of our method are as follows:
\begin{itemize}
\item {\bfseries{Adaptive Missing-Modality Branch Selector}}: A lightweight classifier is designed to dynamically detect the input modality type and activate the reconstruction path for the missing modality in real time.
\item{\bfseries{Athological Information Decoupling Encoder}}: This module decouples shared and modality-specific features from bimodal inputs, enabling deep integration of histological structure information from H\&E images and protein expression patterns from IHC images.
\item{\bfseries{Bidirectional Cross-Modal Reconstruction (CM-GAN)}}: A context-aware generative adversarial network that reconstructs missing modalities within the feature space using cross-modal knowledge transfer.
\item{\bfseries{Modality-Sensitive Feature Attention Module}}: A channel-wise attention mechanism that adaptively adjusts feature weights based on the quality of input modalities, enhancing the model's robustness to data variability.
\end{itemize}

\section{Related Work}

\textbf{Automated HER2 Scoring Based on a Single Modality}:
Given the limitations of subjective manual assessment, researchers have begun exploring automated HER2 evaluation based on single data modalities. An early study by Masmoudi et al. \cite{ref7}developed an image analysis algorithm to objectively quantify HER2 expression levels displayed in IHC images—one of the first attempts at computer-aided prediction of HER2 expression. Most prior deep learning efforts have focused on single image modalities. Several studies targeted IHC slide analysis: Che et al.\cite{ref10} trained convolutional neural networks to identify HER2-positive and -negative tumor cells in whole-slide IHC images, achieving high consistency with pathologists. Xiong et al. \cite{ref8} proposed a comprehensive AI system for HER2 scoring on IHC slides, enhancing inter-slide and inter-laboratory consistency. Similarly, Chauhan et al.\cite{ref9} introduced a deep learning approach that leveraged multi-resolution features from IHC slides to improve HER2 scoring, demonstrating that combining features from both high and low magnifications enhances classification performance for 0/1+, 2+, and 3+ cases. These IHC-based methods benefit from the staining’s specificity but fail to utilize morphological context available in H\&E staining.

On the other hand, some studies have attempted to predict HER2 expression solely from H\&E images by extracting morphological cues. Farahmand et al. \cite{ref11} trained deep learning models on tumor regions in H\&E-stained slides to predict HER2 positivity and response to HER2-targeted therapies. Notably, their H\&E-based model achieved an AUC of approximately 0.88 in identifying HER2-positive tumors, but its performance was inherently limited due to the absence of direct protein expression information. Rasmussen et al. \cite{ref12} focused on the ambiguous HER2 2+ cases in IHC images and developed a model that used H\&E image features to predict the final HER2 status of 2+ cases as a decision-support tool. Although H\&E-based models are appealing due to the widespread use of H\&E slides, their performance often lags behind IHC-based approaches because H\&E lacks intrinsic HER2 expression signals. Overall, whether using H\&E or IHC, single-modality approaches face performance bottlenecks, as each modality reflects only part of the underlying biological information\cite{ref12} This limitation has prompted efforts to leverage complementary information via multimodal fusion.

\textbf{Applications of Multimodal Learning in Breast Cancer Diagnosis}:
Multimodal data fusion has shown great promise in various medical imaging tasks, as integrating heterogeneous data allows for more comprehensive disease characterization \cite{ref17}. In breast cancer diagnosis, examples beyond pathology already exist. For instance, McKinney et al.\cite{ref18} combined multi-view imaging and clinical insights to improve cancer detection, while Joo et al. \cite{ref19} fused imaging and clinical variables for outcome prediction. However, in pathology-based HER2 assessment, multimodal learning remains in its early stages. The study by Liu et al. \cite{ref14} represents a notable effort to bridge H\&E and IHC: they constructed the Breast Cancer Immunohistochemistry (BCI) dataset, which includes 4,870 paired H\&E and IHC image patches. They proposed a pyramid Pix2Pix generative model to translate H\&E images into IHC counterparts, demonstrating the feasibility of cross-modal staining synthesis for HER2 expression prediction. While their focus was image generation (enhancing or predicting IHC appearance from H\&E) rather than direct HER2 scoring, it laid the groundwork for cross-modal approaches. To our knowledge, no prior published methods have yet jointly fused H\&E and IHC modalities in an end-to-end framework for HER2 grading.

\textbf{Missing Modalities and Dynamic Fusion Strategies}:
A practical challenge in multimodal pathology is the potential absence or unavailability of one modality. Simple imputation methods—such as filling missing inputs with zeros or mean values—are commonly used but often degrade performance due to the introduction of unnatural features. In general machine learning literature, handling missing modalities involves learning robust representations. For example, Wang et al. \cite{ref15} proposed a shared-specific feature modeling framework that learns latent features shared across all modalities and modality-specific features, enabling prediction even when one modality is missing. Their approach, applied in visual tasks, inspired us to adopt decoupled shared and specific feature extractors. However, such techniques have yet to be applied in pathology within intelligent medical systems. Another line of work leverages generative models to infer missing modalities. Our use of a Pix2Pix-based Generative Adversarial Network (GAN) aligns with this idea: by generating pseudo-IHC images from H\&E inputs (or vice versa), we effectively impute missing information via data-driven inference rather than static imputation. This approach resembles data augmentation across modalities and captures the complex nonlinear relationships better than simpler methods like linear interpolation.

Equally important in multimodal fusion is the strategy for integrating information across modalities. Many existing multimodal models rely on simple concatenation or fixed fusion rules, treating each modality’s contribution as static. However, such strategies are suboptimal when modality relevance varies case-by-case. Attention mechanisms have emerged as powerful tools for adaptive fusion. The Convolutional Block Attention Module (CBAM) proposed by Woo et al. \cite{ref16} is a representative technique that recalibrates feature importance via channel and spatial attention in CNNs. Although originally designed for single-modality images, we were inspired by its mechanism and introduced it into multimodal fusion to dynamically adjust the weight of each modality’s features. For example, when IHC images are blurry or noisy, the model can automatically down-weight their contribution and rely more on H\&E features, and vice versa. This mechanism enables content-aware reliability assessment and adaptive weighting of modalities, enhancing fusion performance.

To our knowledge, this study is the first in medical multimodal tasks to integrate dynamic attention fusion with architectures designed to handle missing modalities. Our proposed framework combines decoupled pathological information encoders, a bidirectional cross-modal reconstruction module, and an attention-based adaptive fusion strategy. It fully integrates multimodal information when available and maintains robust performance when modalities are missing or of low quality. This design effectively overcomes the limitations of existing approaches in dealing with modality heterogeneity and incompleteness.
\section{Methods}

This study developed a deep learning framework for predicting HER2 expression status, capable of handling incomplete histopathology (HE) and immunohistochemistry (IHC) inputs. The core architecture of the framework is illustrated in Figure 1.
\begin{figure}[htbp]
  \centering
  \includegraphics[width=\linewidth]{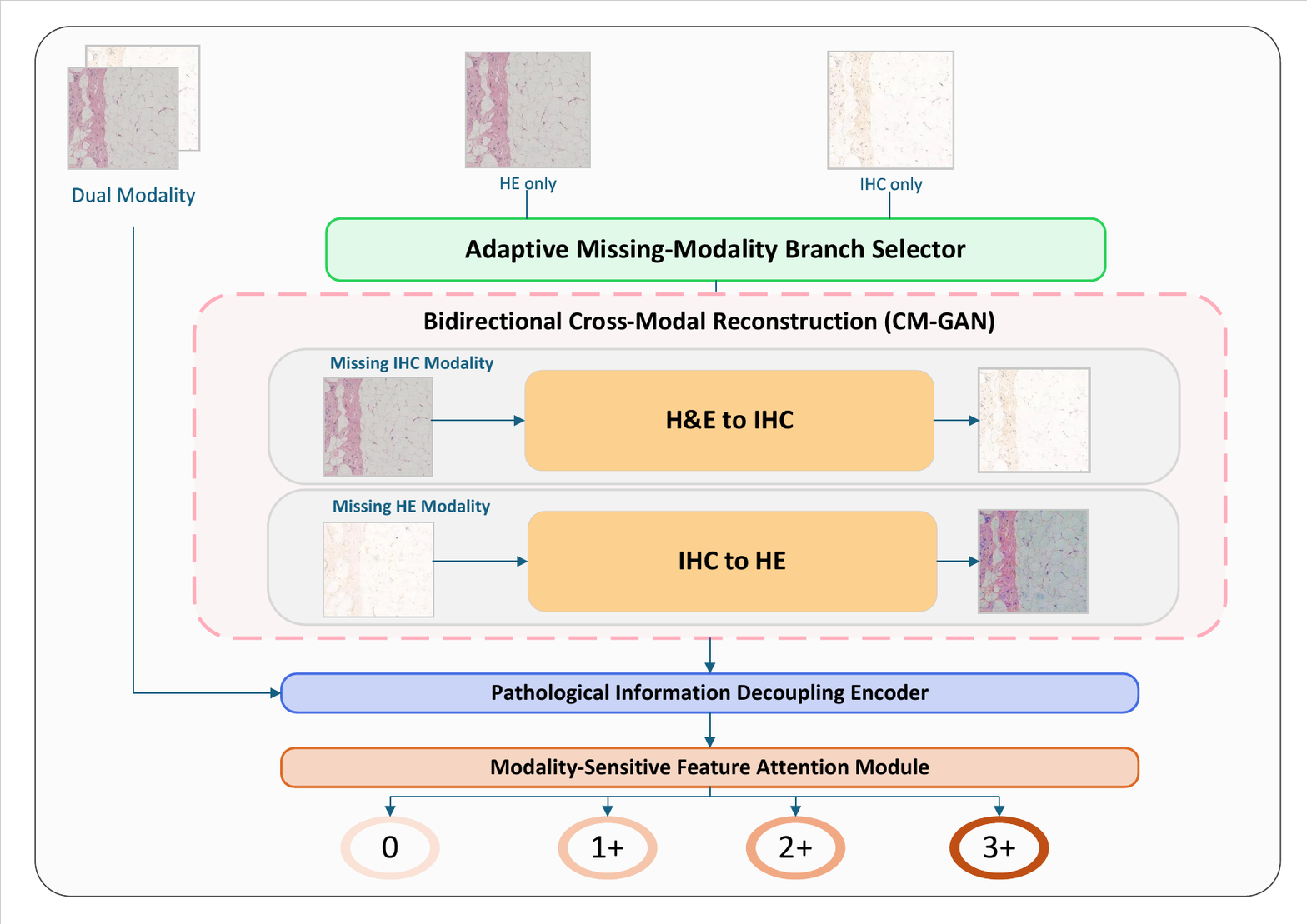}
  \caption{The main architecture of this framework includes a missing modality branch selector, a pathological information decoupling encoder, a bidirectional cross-modal reconstruction network, and a modality-sensitive feature attention module, among others. }
\end{figure}

\subsection{Adaptive Missing-Modality Branch Selector}

The missing modality branch selector is one of the key innovations that enables flexible inference in this framework. By leveraging a dual-branch architecture and a dynamic path selection mechanism, it establishes an adaptive processing workflow tailored to real-world clinical scenarios, enabling intelligent adaptation to both the present and missing modalities.

The system relies on prior manual judgment to determine whether the current input is single-modality (H\&E or IHC) or dual-modality (H\&E + IHC). If the input is dual-modality, it is directly passed to the pathology information disentanglement encoder for multimodal feature fusion and prediction. For inputs identified as single-modality, the model further utilizes a modality classification module based on a GoogleNet architecture to determine the specific modality. This classifier first extracts spatial statistical features of the image using a global average pooling layer, followed by a fully connected layer to identify whether the input is H\&E or IHC. After classification, the corresponding image is forwarded to the dynamic bi-directional reconstruction module, CM-GAN, to generate the missing modality.

\subsection{Bidirectional Cross-Modal Reconstruction Network}

This module adopts a dynamic bi-directional cross-modal reconstruction network (CM-GAN) to achieve semantically consistent completion of missing modalities. For example, when only an H\&E image is provided, the system invokes the HE→IHC reconstruction model; conversely, when only an IHC image is available, the IHC→HE reconstruction model is activated.

As illustrated in Figure 2 and inspired by the work of Liu et al. \cite{ref14},This module employs a dynamic bi-directional cross-modal reconstruction module (CM-GAN) to achieve semantically consistent completion of missing modalities. For example, when only an H\&E image is provided, the system invokes the HE→IHC reconstruction model; conversely, when only an IHC image is available, the IHC→HE reconstruction model is activated.

Inspired by the work of Liu et al. \cite{ref14}, when the input consists of a single-modality image (H\&E or IHC), we adopt a Pyramid Pix2pix architecture to reconstruct the corresponding missing modality image from the single-modality feature space, as illustrated in Figure 2 (only the H\&E→IHC reconstruction is shown here; the reverse process follows the same principle). Specifically, the model first extracts hierarchical features of the single-modality image using multi-scale residual blocks, followed by the application of a spatial attention mechanism to locate key regions. Then, skip connections are used to fuse low-level details with high-level semantic information, enabling the reconstruction of the missing modality image. Finally, the reconstructed image is combined with the original modality image and fed into the subsequent pathology information disentanglement encoder for feature fusion.

\begin{figure}[htbp]
  \centering
  \includegraphics[width=\linewidth]{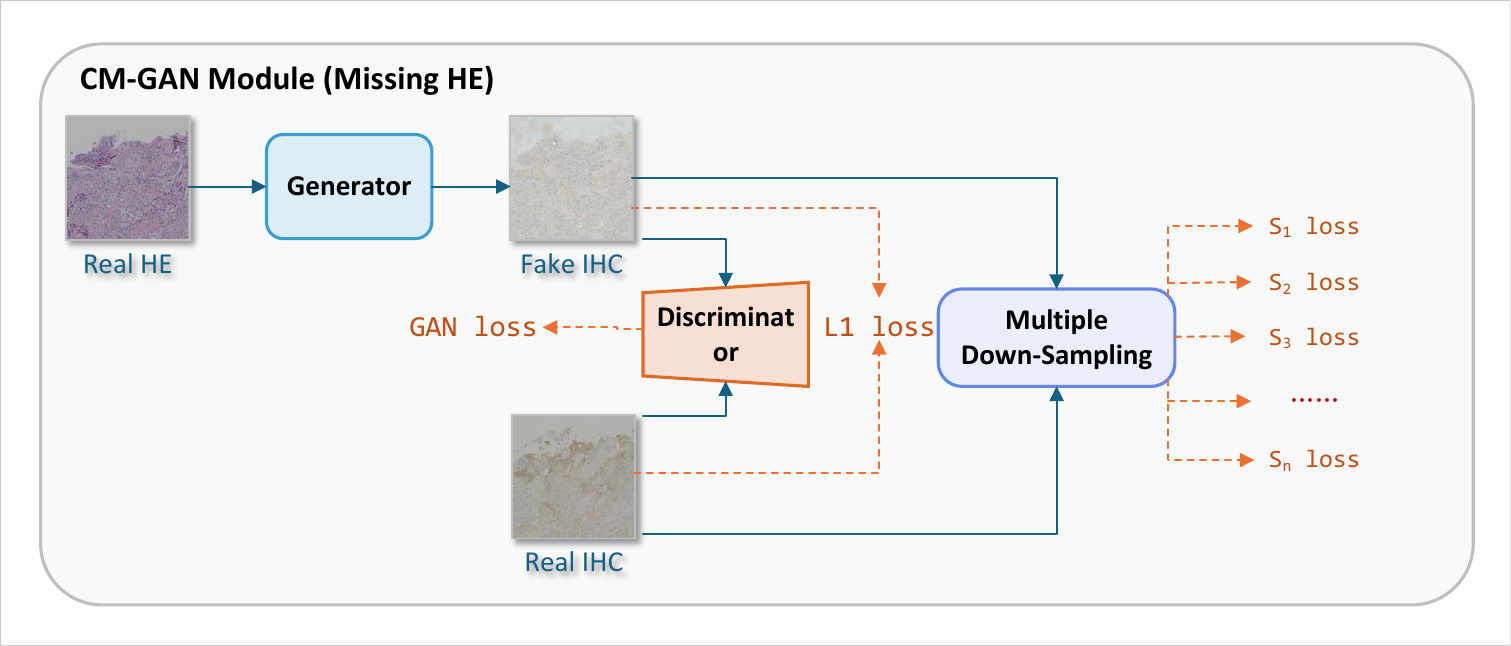}
  \caption{Structure diagram of bidirectional cross-modal reconstruction module.}
\end{figure}

\subsection{Pathological Information Decoupling Encoder}

To integrate the morphological characteristics of H\&E images and the protein expression patterns of IHC images from a clinical interpretability perspective, we draw inspiration from the Share-Specific Encoder design proposed by Wang et al. \cite{ref15} and implement a pathology information disentanglement encoder. This encoder is designed to decouple and synergize cross-modal shared and modality-specific features.

As shown in Figure 3, the module employs a pre-trained ResNet50 as the shared encoder to extract cross-modal shared features (denoted as $F_s$) from both H\&E and IHC images, thereby capturing the common patterns underlying tissue morphology and molecular expression.

Meanwhile, to retain modality-specific information, the module introduces two independent encoder branches dedicated to extracting modality-specific features: texture and topological features ($F_{he}$) from H\&E images, and protein distribution features ($F_{ihc}$) from IHC images.

To further enhance feature disentanglement, the module adopts the joint optimization strategy proposed by Wang et al.\cite{ref15}, which incorporates both Domain Classification Loss (DCO) and Distribution Alignment Loss (DAO) during training. DCO enforces the specificity of modality features via binary cross-entropy loss, encouraging them to discriminate between H\&E and IHC modalities. DAO, based on Maximum Mean Discrepancy (MMD), constrains the distributions of shared features across modalities to align within a unified feature space.

\begin{figure}[htbp]
  \centering
  \includegraphics[width=\linewidth]{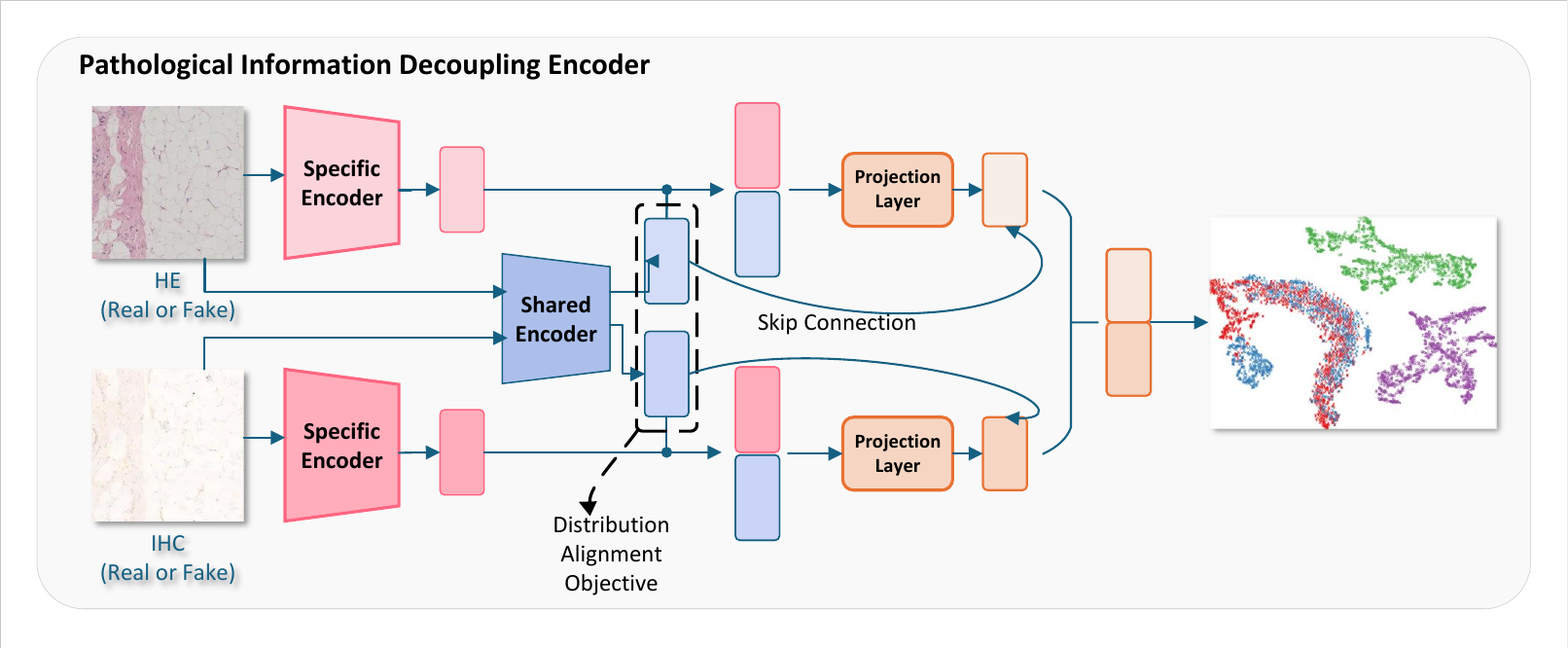}
  \caption{Structure diagram of pathological information decoupling encoder}
\end{figure}

\subsection{Modality-Sensitive Feature Attention Module}

To enable a dynamic weighting strategy based on the quality of different modality images, this study introduces a modality-sensitive feature attention module designed for adaptive fusion of multimodal information. Inspired by the Convolutional Block Attention Module (CBAM) proposed by Woo et al.\cite{ref16}, this module incorporates image quality assessment logic to automatically evaluate and adjust the importance weights of each modality’s features, thereby enhancing the discriminative power of the fused representation.

Specifically, when a certain modality (e.g., IHC) suffers from issues such as weak staining or high background noise, the model can automatically increase the contribution of another modality (e.g., H\&E), enabling more robust cross-modal joint modeling.

To achieve this, the module first concatenates the shared features $F_s$ with the modality-specific features $F_{he}$ and $F_{ihc}$ along the channel dimension, forming a unified multimodal feature sequence. Considering the potentially drastic increase in dimensionality after concatenation, a 1×1 convolution is applied to reduce the number of channels, thereby lowering computational complexity while preserving core semantic information.

Next, a channel attention mechanism is introduced to model the importance of different channels. This mechanism utilizes both global average pooling and max pooling, feeding the resulting features into a multi-layer perceptron (MLP), and applying a Sigmoid activation function to output the channel-wise attention weights. The computation of this attention is as follows:

\begin{equation}
\begin{aligned}
M_c(F) &= \sigma\left(\text{MLP}\left(\text{AvgPool}(F)\right) + \text{MLP}\left(\text{MaxPool}(F)\right)\right) \\
       &= \sigma\left(W_1 \left(W_0(F_{\text{avg}}^c)\right) + W_1(F_{\text{max}}^c)\right)
\end{aligned}
\end{equation}

where $F_{\text{avg}}^c$ and $F_{\text{max}}^c$ represent the average-pooled and max-pooled results across the channel dimension, respectively. $W_0$ and $W_1$ are the MLP weight parameters, and $\sigma(\cdot)$ denotes the Sigmoid function, used to produce a normalized distribution of channel weights.

Finally, the module applies the computed channel attention weights to the concatenated features for weighted fusion, producing the final fused features $F_{\text{fused}}$. These are then passed to the subsequent classification head to accurately predict HER2 expression levels.

\begin{figure}[htbp]
    \centering
    \includegraphics[width=1\linewidth]{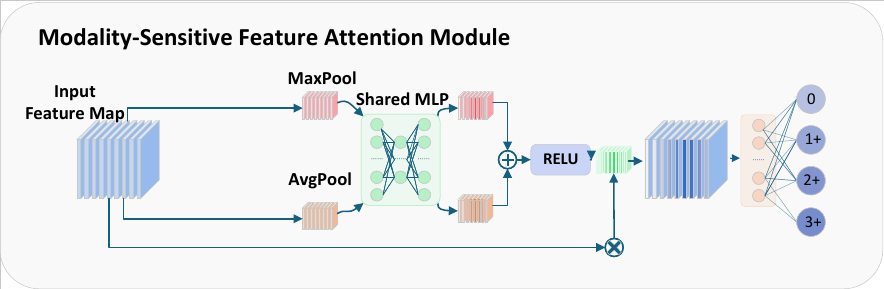}
    \caption{Structure diagram of modality-sensitive feature attention module.}
    \label{fig:enter-label}
\end{figure}

\subsection{Loss Function}

In this study, we consider the task characteristics of three key modules during training: the missing modality branch selector, the pathological information disentanglement encoder, and the dynamic bidirectional cross-modal reconstruction module. Each module has different optimization objectives, and thus different loss functions are applied and jointly trained to improve overall model performance and robustness.

\textbf{The loss of missing-modality branch selector and final classifier head.}This module is designed for the classification task, and a weighted cross-entropy loss is adopted:

\begin{equation}
\begin{aligned}
\mathcal{L}_{\text{cls}} = - \sum_{i=1}^{C} w_i \cdot y_i \cdot \log(\hat{y}_i)
\end{aligned}
\end{equation}

where $C$ represents the four HER2 expression levels (0, 1+, 2+, 3+); $y_i$ is the one-hot encoding of the ground truth label; $\hat{y}_i$ denotes the predicted probability output of the model; and $w_i$ is the class-specific weight to address class imbalance.

\textbf{The loss of pathological information decoupling encoder.}
To enhance the domain-invariant modeling ability of pathological representations across different modalities, we follow the structure proposed by Wang et al.\cite{ref15} and introduce a domain classification loss and a distribution alignment loss in this module. The total loss is defined as:

\begin{equation}
\begin{aligned}
\mathcal{L}_{\text{enc}} = \lambda_1 \cdot \mathcal{L}_{\text{domain}} + \lambda_2 \cdot \mathcal{L}_{\text{align}}
\end{aligned}
\end{equation}

where $\mathcal{L}_{\text{domain}}$ is a cross-entropy loss for distinguishing modality domains, encouraging the encoder to learn domain-invariant features; $\mathcal{L}_{\text{align}}$ aligns feature distributions via KL divergence or Maximum Mean Discrepancy (MMD); $\lambda_1$ and $\lambda_2$ are the weighting factors for the two sub-losses.

\textbf{The loss of dynamic bidirectional cross-modal reconstruction.}This module facilitates information complementation and reconstruction across modalities. Inspired by the Pyramid Pix2Pix framework proposed by Liu et al.\cite{ref14}, we employ a combination of generative adversarial loss and multi-scale L1 reconstruction loss within a pyramid architecture:

\begin{equation}
\begin{aligned}
\mathcal{L}_{\text{recon}} = \lambda_3 \cdot \mathcal{L}_{\text{GAN}} + \lambda_4 \cdot \sum_{s=1}^{S} \mathcal{L}_{\text{L1}}^{(s)}
\end{aligned}
\end{equation}

where $\mathcal{L}_{\text{GAN}}$ is the standard adversarial loss optimizing the generator and discriminator;

\begin{equation}
\begin{aligned}
\mathcal{L}_{\text{L1}}^{(s)} = \| G^{(s)}(z) - x^{(s)} \|_1
\end{aligned}
\end{equation}

$\mathcal{L}^{(s)}_{\text{L1}}$ is the L1 reconstruction error at the $s$-th level of the pyramid; $G^{(s)}(z)$ is the generated image at scale $s$; $x^{(s)}$ is the corresponding ground truth pyramid representation; $\lambda_3$ and $\lambda_4$ control the contribution of the GAN and reconstruction losses, respectively; $S$ is the total number of pyramid levels.

\textbf{Overall Loss.}The overall loss function is the summation of all module-specific loss terms:

\begin{equation}
\begin{aligned}
\mathcal{L}_{\text{total}} = \mathcal{L}_{\text{cls}} + \mathcal{L}_{\text{enc}} + \mathcal{L}_{\text{recon}}
\end{aligned}
\end{equation}

By jointly optimizing these multi-objective losses, the model achieves improved classification accuracy, robust cross-modal feature modeling, and reliable performance under missing modality conditions.

\section{Experiments}

To comprehensively validate the effectiveness and robustness of the proposed model, a series of systematic experiments were conducted in this study, including: (1) dataset preparation and evaluation metric definition; (2) standardized implementation details to ensure experimental reproducibility; (3) comparative experiments demonstrating that cross-modal reconstruction enhancement and advanced fusion strategies outperform traditional unimodal inputs and simple concatenation methods; and (4) ablation studies analyzing the contribution of the modality-sensitive feature attention module to overall performance. Experimental results show that the model maintains high prediction accuracy and robustness even in the presence of missing modalities or low-quality inputs, highlighting its strong potential for multimodal HER2 expression prediction tasks.

\subsection{Dataset and Evaluation Metrics}

\textbf{Dataset.}This study employs the BCI Dataset (refer to Liu et al., 2018 \cite{ref14}), which contains 4,870 pairs of strictly matched H\&E and IHC whole-slide images. Each image pair is explicitly annotated with HER2 expression levels (0, 1+, 2+, or 3+). The dataset holds important clinical value, covering the full range of HER2 expression from negative to strongly positive. The images accurately reflect both the morphological characteristics of pathological tissues and HER2 protein expression, making the dataset highly suitable for multimodal fusion research and experiments simulating modality dropout.

\textbf{Data Preprocessing.}To standardize input image dimensions and enhance the model’s generalization ability, all images underwent data augmentation and normalization. The data augmentation techniques included random rotation (±15°), horizontal and vertical flipping, random cropping, and random brightness and contrast adjustment. Normalization was mainly used to unify image resolution. Additionally, HER2 grading information embedded in the filenames was used to convert the labels into one-hot encoded vectors. The dataset was split into training, validation, and test sets in a ratio of 8:1:1.

\textbf{Performance Metrics.}To comprehensively evaluate the classification performance of the model, we used a series of metrics, including accuracy, recall, precision, F1-score, and confusion matrix analysis. Among them, the F1-score provides a balanced measure of the model’s sensitivity to class imbalance and its overall robustness. Furthermore, we used t-SNE visualization of the feature space to improve model interpretability, offering an intuitive understanding of how the model utilizes multimodal features and assigns weights during the prediction process.

\begin{figure}[htbp]
    \centering
    \includegraphics[width=1\linewidth]{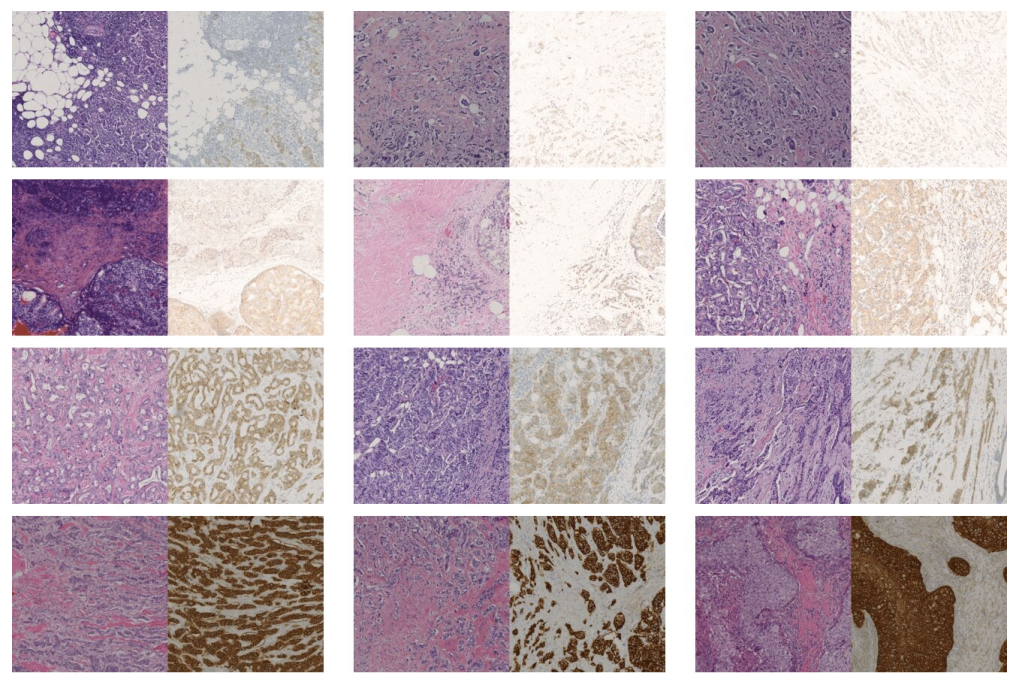}
    \caption{Examples of H\&E-IHC image pairs corresponding to the four HER2 expression levels (0, 1+, 2+, 3+ from top to bottom). In each row, the left and right images represent H\&E staining (left) and IHC staining (right), respectively, illustrating the morphological and staining characteristics at different expression levels.}
    \label{fig:enter-label}
\end{figure}

\subsection{Implementation Details}

To ensure the fairness and reproducibility of the experimental results, all experiments were conducted under a standardized hardware and software environment. The hardware setup included a server equipped with an Intel® Xeon® Platinum 8352V CPU (2.10GHz) and an NVIDIA RTX 4090 GPU with 24GB of memory. On the software side, the environment was built on Ubuntu 18.04, using Python 3.8 and PyTorch 2.4 as the core programming language and framework. During model training, the AdamW optimizer was employed with an initial learning rate of 1e-4. To prevent overfitting, a polynomial learning rate decay strategy was adopted.

\subsection{Comparative Experiments}

This section systematically validates the effectiveness of the proposed method through two sets of controlled experiments. All experiments were conducted under a unified data split and consistent hyperparameter settings to ensure the comparability of results.

\textbf{Unimodal vs. Cross-Modal Reconstruction Enhancement.}
In the unimodal baseline method, only real H\&E or IHC images were used as input, with MobileNetV2 serving as the backbone network. In contrast, the cross-modal reconstruction enhancement method employs a pyramid pix2pix framework to generate cross-modal images (e.g., Fake IHC or Fake HE). The real modality and the reconstructed modality are jointly used as inputs to the model, and feature-level fusion is achieved through a shared-specific feature extraction module. As shown in Table 1, the cross-modal reconstruction method significantly outperforms the unimodal baseline. For instance, the Real H\&E + Fake IHC combination achieved an accuracy of 94.25\%, representing a 22.81\% improvement over the HE unimodal baseline (71.44\%), and the F1-score increased to 0.9609. Similarly, the Real IHC + Fake H\&E combination achieved an accuracy of 90.28\%, which is 12.90\% higher than the IHC unimodal baseline (77.38\%). These results indicate that cross-modal generation can effectively compensate for the limitations of single-modal information, thereby enhancing the generalization ability of the model.

\begin{table*}[htbp]
  \caption{Performance Comparison between Unimodal and Cross-Modal Reconstruction Methods.}
  \label{tab:comparison}
  \centering
  \begin{tabular}{lccc}
    \toprule
    \textbf{Method Type} & \textbf{Accuracy} & \textbf{F1-score} & \textbf{Improvement (vs. Unimodal)} \\
    \midrule
    H\&E Unimodal & 71.44\% & 0.6697 & - \\
    IHC Unimodal & 77.38\% & 0.7646 & - \\
    Real HE + Fake IHC (ours) & 94.25\% & 0.9609 & +22.81 (vs. H\&E Unimodal) \\
    Real IHC + Fake H\&E (ours) & 90.28\% & 0.9251 & +12.90 (vs. IHC Unimodal) \\
    \bottomrule
  \end{tabular}
\end{table*}

\textbf{Dual-Modal Simple Concatenation vs. Pathological Information Decoupling Encoding.}The dual-modal concatenation baseline fuses real H\&E and real IHC images through a simple concatenation operation, employing the same backbone network as used in the unimodal setting. In contrast, the pathological information decoupling encoding method jointly inputs real H\&E and real IHC images, and achieves more refined feature interaction by decoupling modality-shared and modality-specific features. As shown in Table 2, the feature fusion method achieves an accuracy of 95.09\%, representing a 4.29\% improvement over the concatenation baseline (92.00\%), and the F1-score increases to 0.9532. This indicates that simple concatenation only enables shallow feature interaction, whereas the pathological information decoupling encoding architecture can effectively capture cross-modal complementary information and suppress noise interference between modalities.

\begin{table*}[htbp]
  \caption{Performance Comparison of Dual-Modal Fusion Strategies.}
  \label{tab:dual_fusion}
  \centering
  \begin{tabular}{lccc}
    \toprule
    \textbf{Input Type} & \textbf{Accuracy} & \textbf{F1-score} & \textbf{Improvement (vs. Concatenation Baseline)} \\
    \midrule
    simple concat & 92.00\% & 0.9103 & - \\
    Real HE + Real IHC (ours) & 95.09\% & 0.9532 & +4.29 (vs. simple concat) \\
    \bottomrule
  \end{tabular}
\end{table*}

\subsection{Ablation Studies}

The ablation experiments validated the contribution of the key module to the overall model performance. In these experiments, the model without the modality-sensitive feature attention module was compared with the full model incorporating this module. In the baseline model, shared and specific features were directly concatenated and then reduced in dimension using a 1×1 convolution. In contrast, the full model introduced the modality-sensitive feature attention module, which dynamically adjusts the weights of different modalities in real time. Experimental results showed that the full model achieved an accuracy of 95.32\% and an F1-score of 0.9532, representing a 4.91\% improvement in F1-score compared to the baseline model's 92.41\% accuracy and 0.9041 F1-score. These findings clearly demonstrate that this module significantly enhances the effectiveness of feature fusion, thereby improving the model's overall predictive performance under conditions of poor modality reconstruction quality.

\begin{table}[htbp]
  \caption{Ablation Experiment Results for Modality-Sensitive Feature Attention Module.}
  \label{tab:CAM_removal_ablation}
  \centering
  \resizebox{1.0\linewidth}{!}{
    \begin{tabular}{cccc}
      \toprule
      \textbf{Attention Inclusion} & \textbf{Accuracy} & \textbf{F1-score} & \textbf{Improvement} \\
      \midrule
      $\surd$ & 95.32\% & 0.9532 & +4.91 \\
             & 92.41\% & 0.9041 & - \\
      \bottomrule
    \end{tabular}
  }
\end{table}

\subsection{Component Evaluation}

\textbf{Missing Modality Branch Selector.} To dynamically adapt to unimodal or multimodal input scenarios, a lightweight branch selector module is introduced to determine whether a modality is missing. This module is trained as a binary classifier and achieves a classification accuracy of 99.95\%, recall of 99.95\%, and an F1-score of 0.9995, indicating its near-perfect ability to distinguish between complete and incomplete modality inputs.

\textbf{Bidirectional Cross-Modal Reconstruction Module.} To verify the quality of reconstructed modality inputs, we quantitatively evaluate the cross-modal generation results using PSNR and SSIM. The H\&E-to-IHC reconstruction achieved a PSNR of 18.48 dB and SSIM of 0.51, while the IHC-to-H\&E reconstruction yielded a PSNR of 17.24 dB and SSIM of 0.39. These values demonstrate that the cross-modal generator can produce structurally relevant and visually consistent images, providing effective complementary information for downstream classification tasks.

\begin{figure}
    \centering
    \includegraphics[width=1\linewidth]{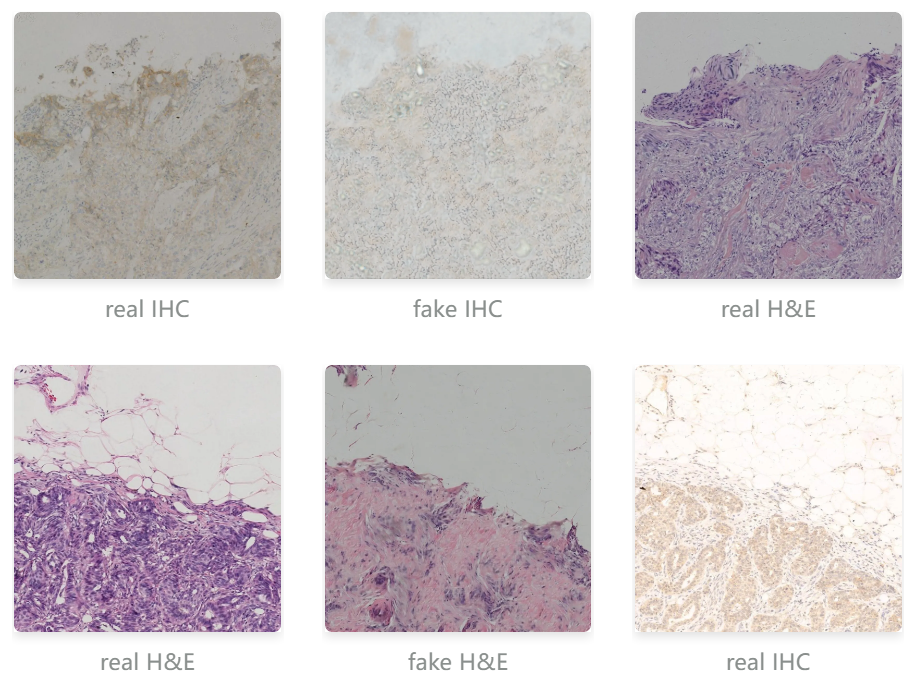}
    \caption{Cross-modal reconstruction comparison. Real IHC/H\&E images and model-generated fake IHC/H\&E images are shown, verifying the bidirectional reconstruction module’s ability to compensate for missing modalities, with structural consistency between fake and real images.}
    \label{fig:enter-label}
\end{figure}

\section{Conclusion}
The dual-modal optional input model proposed in this study
leverages dynamic branch selection and bidirectional cross-modal
reconstruction mechanisms to achieve flexible HER2 prediction
under both unimodal and dual-modal conditions, effectively
overcoming the rigid dependency of traditional methods on
complete modality inputs. Notably, the model attains an accuracy
of 95.09\% with dual-modal input and maintains a high
performance of 94.45\% even with solely HE input, thereby
significantly reducing clinical reliance on IHC staining
equipment.

A key limitation of this study is that all data were sourced from
a single institution. Although data augmentation was employed to
simulate variability, the model has not yet been evaluated on
external datasets collected from diverse institutions or scanners. In future work, we plan to conduct multicenter evaluations to verify the model's generalizability across different staining protocols and imaging devices. In such cases, maintaining high performance may require integrating domain adaptation strategies or fine-tuning the model on new data. Furthermore, we aim to extend this dynamic multimodal approach to other pathological tasks. Future studies could focus on including combining H\&E images with genomic data or radiological images to predict outcomes, or integrating multiple immunohistochemical markers in a single model. The core concept of "activation on demand" is expected to be very useful in applications with inconsistent data availability.

\section{Acknowledgments}
This work was financially supported by Natural Science Foundation of China (grant number 62271466).

\bibliographystyle{ACM-Reference-Format}
\balance
\bibliography{sample-base}


\begin{thebibliography}{19}

%%% the Web address.

\ifx \showCODEN    \undefined \def \showCODEN     #1{\unskip}     \fi
\ifx \showISBNx    \undefined \def \showISBNx     #1{\unskip}     \fi
\ifx \showISBNxiii \undefined \def \showISBNxiii  #1{\unskip}     \fi
\ifx \showISSN     \undefined \def \showISSN      #1{\unskip}     \fi
\ifx \showLCCN     \undefined \def \showLCCN      #1{\unskip}     \fi
\ifx \shownote     \undefined \def \shownote      #1{#1}          \fi
\ifx \showarticletitle \undefined \def \showarticletitle #1{#1}   \fi
\ifx \showURL      \undefined \def \showURL       {\relax}        \fi
% invisible to TeX
\providecommand\bibfield[2]{#2}
\providecommand\bibinfo[2]{#2}
\providecommand\natexlab[1]{#1}
\providecommand\showeprint[2][]{arXiv:#2}

\bibitem[Ahn et~al\mbox{.}(2020)]%
        {ref3}
\bibfield{author}{\bibinfo{person}{S. Ahn}, \bibinfo{person}{J.~W. Woo}, \bibinfo{person}{K. Lee}, {and} \bibinfo{person}{S.~Y. Park}.} \bibinfo{year}{2020}\natexlab{}.
\newblock \showarticletitle{HER2 Status in Breast Cancer: Changes in Guidelines and Complicating Factors for Interpretation}.
\newblock \bibinfo{journal}{\emph{Journal of Pathology and Translational Medicine}} \bibinfo{volume}{54}, \bibinfo{number}{1} (\bibinfo{year}{2020}), \bibinfo{pages}{34--44}.
\newblock


\bibitem[Chauhan et~al\mbox{.}(2025)]%
        {ref9}
\bibfield{author}{\bibinfo{person}{E. Chauhan}, \bibinfo{person}{A. Sharma}, \bibinfo{person}{A. Sharma}, \bibinfo{person}{V. Nishadham}, \bibinfo{person}{A. Ghughtyal}, \bibinfo{person}{A. Kumar}, {and} \bibinfo{person}{et al.}} \bibinfo{year}{2025}\natexlab{}.
\newblock \showarticletitle{Contrasting Low and High-Resolution Features for HER2 Scoring using Deep Learning}.
\newblock \bibinfo{journal}{\emph{arXiv preprint}}  \bibinfo{volume}{arXiv:2503.22069} (\bibinfo{year}{2025}).
\newblock


\bibitem[Che et~al\mbox{.}(2023)]%
        {ref10}
\bibfield{author}{\bibinfo{person}{Y. Che}, \bibinfo{person}{F. Ren}, \bibinfo{person}{X. Zhang}, \bibinfo{person}{L. Cui}, \bibinfo{person}{H. Wu}, {and} \bibinfo{person}{Z. Zhao}.} \bibinfo{year}{2023}\natexlab{}.
\newblock \showarticletitle{Immunohistochemical HER2 Recognition and Analysis of Breast Cancer Based on Deep Learning}.
\newblock \bibinfo{journal}{\emph{Diagnostics}} \bibinfo{volume}{13}, \bibinfo{number}{2} (\bibinfo{year}{2023}), \bibinfo{pages}{263}.
\newblock


\bibitem[Farahmand et~al\mbox{.}(2022)]%
        {ref11}
\bibfield{author}{\bibinfo{person}{S. Farahmand}, \bibinfo{person}{A.~I. Fernandez}, \bibinfo{person}{F.~S. Ahmed}, \bibinfo{person}{D.~L. Rimm}, \bibinfo{person}{J.~H. Chuang}, \bibinfo{person}{E. Reisenbichler}, {and} \bibinfo{person}{K. Zarringhalam}.} \bibinfo{year}{2022}\natexlab{}.
\newblock \showarticletitle{Deep Learning Trained on Hematoxylin and Eosin Tumor Region‑of‑Interest Predicts HER2 Status and Trastuzumab Treatment Response in HER2+ Breast Cancer}.
\newblock \bibinfo{journal}{\emph{Modern Pathology}} \bibinfo{volume}{35}, \bibinfo{number}{1} (\bibinfo{year}{2022}), \bibinfo{pages}{44--51}.
\newblock


\bibitem[Furrer et~al\mbox{.}(2015)]%
        {ref6}
\bibfield{author}{\bibinfo{person}{D. Furrer}, \bibinfo{person}{F. Sanschagrin}, \bibinfo{person}{S. Jacob}, {and} \bibinfo{person}{C. Diorio}.} \bibinfo{year}{2015}\natexlab{}.
\newblock \showarticletitle{Advantages and Disadvantages of Technologies for HER2 Testing in Breast Cancer Specimens}.
\newblock \bibinfo{journal}{\emph{American Journal of Clinical Pathology}} \bibinfo{volume}{144}, \bibinfo{number}{5} (\bibinfo{year}{2015}), \bibinfo{pages}{686--703}.
\newblock


\bibitem[Joo et~al\mbox{.}(2021)]%
        {ref19}
\bibfield{author}{\bibinfo{person}{S. Joo}, \bibinfo{person}{E.~S. Ko}, \bibinfo{person}{S. Kwon}, \bibinfo{person}{E. Jeon}, \bibinfo{person}{H. Jung}, \bibinfo{person}{J.~Y. Kim}, {and} \bibinfo{person}{et al.}} \bibinfo{year}{2021}\natexlab{}.
\newblock \showarticletitle{Multimodal Deep Learning Models for the Prediction of Pathologic Response to Neoadjuvant Chemotherapy in Breast Cancer}.
\newblock \bibinfo{journal}{\emph{Scientific Reports}} \bibinfo{volume}{11}, \bibinfo{number}{1} (\bibinfo{year}{2021}), \bibinfo{pages}{18800}.
\newblock


\bibitem[Liu et~al\mbox{.}(2022)]%
        {ref14}
\bibfield{author}{\bibinfo{person}{S. Liu}, \bibinfo{person}{C. Zhu}, \bibinfo{person}{F. Xu}, \bibinfo{person}{X. Jia}, \bibinfo{person}{Z. Shi}, {and} \bibinfo{person}{M. Jin}.} \bibinfo{year}{2022}\natexlab{}.
\newblock \showarticletitle{BCI: Breast Cancer Immunohistochemical Image Generation through Pyramid Pix2Pix}. In \bibinfo{booktitle}{\emph{Proceedings of the IEEE/CVF Conference on Computer Vision and Pattern Recognition (CVPR)}}. \bibinfo{pages}{1815--1824}.
\newblock


\bibitem[Masmoudi et~al\mbox{.}(2009)]%
        {ref7}
\bibfield{author}{\bibinfo{person}{H. Masmoudi}, \bibinfo{person}{S.~M. Hewitt}, \bibinfo{person}{N. Petrick}, \bibinfo{person}{K.~J. Myers}, {and} \bibinfo{person}{M.~A. Gavrielides}.} \bibinfo{year}{2009}\natexlab{}.
\newblock \showarticletitle{Automated Quantitative Assessment of HER‑2/neu Immunohistochemical Expression in Breast Cancer}.
\newblock \bibinfo{journal}{\emph{IEEE Transactions on Medical Imaging}} \bibinfo{volume}{28}, \bibinfo{number}{6} (\bibinfo{year}{2009}), \bibinfo{pages}{916--925}.
\newblock


\bibitem[McKinney et~al\mbox{.}(2020)]%
        {ref18}
\bibfield{author}{\bibinfo{person}{S.~M. McKinney}, \bibinfo{person}{M. Sieniek}, \bibinfo{person}{V. Godbole}, \bibinfo{person}{J. Godwin}, \bibinfo{person}{N. Antropova}, \bibinfo{person}{H. Ashrafian}, {and} \bibinfo{person}{et al.}} \bibinfo{year}{2020}\natexlab{}.
\newblock \showarticletitle{International Evaluation of an AI System for Breast Cancer Screening}.
\newblock \bibinfo{journal}{\emph{Nature}} \bibinfo{volume}{577}, \bibinfo{number}{7788} (\bibinfo{year}{2020}), \bibinfo{pages}{89--94}.
\newblock


\bibitem[Nicolò et~al\mbox{.}(2023)]%
        {ref13}
\bibfield{author}{\bibinfo{person}{E. Nicolò}, \bibinfo{person}{L.~Boscolo Bielo}, {and} \bibinfo{person}{et al.}} \bibinfo{year}{2023}\natexlab{}.
\newblock \showarticletitle{The HER2‑low Revolution in Breast Oncology: Steps Forward and Emerging Challenges}.
\newblock \bibinfo{journal}{\emph{Therapeutic Advances in Medical Oncology}}  \bibinfo{volume}{15} (\bibinfo{year}{2023}), \bibinfo{pages}{17588359231152842}.
\newblock


\bibitem[Press et~al\mbox{.}(1994)]%
        {ref5}
\bibfield{author}{\bibinfo{person}{M.~F. Press}, \bibinfo{person}{G. Hung}, \bibinfo{person}{W. Godolphin}, {and} \bibinfo{person}{D.~J. Slamon}.} \bibinfo{year}{1994}\natexlab{}.
\newblock \showarticletitle{Sensitivity of HER‑2/neu Antibodies in Archival Tissue Samples: Potential Source of Error in Immunohistochemical Studies of Oncogene Expression}.
\newblock \bibinfo{journal}{\emph{Cancer Research}} \bibinfo{volume}{54}, \bibinfo{number}{10} (\bibinfo{year}{1994}), \bibinfo{pages}{2771--2777}.
\newblock


\bibitem[Rasmussen et~al\mbox{.}(2022)]%
        {ref12}
\bibfield{author}{\bibinfo{person}{S.~A. Rasmussen}, \bibinfo{person}{V.~J. Taylor}, \bibinfo{person}{A.~P. Surette}, \bibinfo{person}{P.~J. Barnes}, {and} \bibinfo{person}{G.~C. Bethune}.} \bibinfo{year}{2022}\natexlab{}.
\newblock \showarticletitle{Using Deep Learning to Predict Final HER2 Status in Invasive Breast Cancers That Are Equivocal (2+) by Immunohistochemistry}.
\newblock \bibinfo{journal}{\emph{Applied Immunohistochemistry \& Molecular Morphology}} \bibinfo{volume}{30}, \bibinfo{number}{10} (\bibinfo{year}{2022}), \bibinfo{pages}{668--673}.
\newblock


\bibitem[Steyaert et~al\mbox{.}(2023)]%
        {ref17}
\bibfield{author}{\bibinfo{person}{S. Steyaert}, \bibinfo{person}{M. Pizurica}, \bibinfo{person}{D. Nagaraj}, \bibinfo{person}{P. Khandelwal}, \bibinfo{person}{T. Hernandez‑Boussard}, \bibinfo{person}{A.~J. Gentles}, {and} \bibinfo{person}{O. Gevaert}.} \bibinfo{year}{2023}\natexlab{}.
\newblock \showarticletitle{Multimodal Data Fusion for Cancer Biomarker Discovery with Deep Learning}.
\newblock \bibinfo{journal}{\emph{Nature Machine Intelligence}} \bibinfo{volume}{5}, \bibinfo{number}{4} (\bibinfo{year}{2023}), \bibinfo{pages}{351--362}.
\newblock


\bibitem[Sun et~al\mbox{.}(2017)]%
        {ref1}
\bibfield{author}{\bibinfo{person}{Y.~S. Sun}, \bibinfo{person}{Z. Zhao}, \bibinfo{person}{Z.~N. Yang}, \bibinfo{person}{F. Xu}, \bibinfo{person}{H.~J. Lu}, \bibinfo{person}{Z.~Y. Zhu}, {and} \bibinfo{person}{et al.}} \bibinfo{year}{2017}\natexlab{}.
\newblock \showarticletitle{Risk Factors and Preventions of Breast Cancer}.
\newblock \bibinfo{journal}{\emph{International Journal of Biological Sciences}} \bibinfo{volume}{13}, \bibinfo{number}{11} (\bibinfo{year}{2017}), \bibinfo{pages}{1387--1397}.
\newblock


\bibitem[Viale and Bardia(2023)]%
        {ref4}
\bibfield{author}{\bibinfo{person}{Giancarlo Viale} {and} \bibinfo{person}{Aditya Bardia}.} \bibinfo{year}{2023}\natexlab{}.
\newblock \showarticletitle{HER2‑low breast cancer – Diagnostic challenges and opportunities}.
\newblock \bibinfo{journal}{\emph{Targeted Oncology}}  \bibinfo{volume}{18} (\bibinfo{year}{2023}), \bibinfo{pages}{1--9}.
\newblock


\bibitem[Waks and Winer(2019)]%
        {ref2}
\bibfield{author}{\bibinfo{person}{A.~G. Waks} {and} \bibinfo{person}{E.~P. Winer}.} \bibinfo{year}{2019}\natexlab{}.
\newblock \showarticletitle{Breast Cancer Treatment: A Review}.
\newblock \bibinfo{journal}{\emph{JAMA}} \bibinfo{volume}{321}, \bibinfo{number}{3} (\bibinfo{year}{2019}), \bibinfo{pages}{288--300}.
\newblock


\bibitem[Wang et~al\mbox{.}(2023)]%
        {ref15}
\bibfield{author}{\bibinfo{person}{H. Wang}, \bibinfo{person}{Y. Chen}, \bibinfo{person}{C. Ma}, \bibinfo{person}{J. Avery}, \bibinfo{person}{L. Hull}, {and} \bibinfo{person}{G. Carneiro}.} \bibinfo{year}{2023}\natexlab{}.
\newblock \showarticletitle{Multi‑Modal Learning with Missing Modality via Shared‑Specific Feature Modelling}. In \bibinfo{booktitle}{\emph{Proceedings of the IEEE/CVF Conference on Computer Vision and Pattern Recognition (CVPR)}}. \bibinfo{pages}{15878--15887}.
\newblock


\bibitem[Woo et~al\mbox{.}(2018)]%
        {ref16}
\bibfield{author}{\bibinfo{person}{S. Woo}, \bibinfo{person}{J. Park}, \bibinfo{person}{J.~Y. Lee}, {and} \bibinfo{person}{I.~S. Kweon}.} \bibinfo{year}{2018}\natexlab{}.
\newblock \showarticletitle{CBAM: Convolutional Block Attention Module}. In \bibinfo{booktitle}{\emph{Proceedings of the European Conference on Computer Vision (ECCV)}}. \bibinfo{pages}{3--19}.
\newblock


\bibitem[Xiong et~al\mbox{.}(2024)]%
        {ref8}
\bibfield{author}{\bibinfo{person}{Z. Xiong}, \bibinfo{person}{K. Liu}, \bibinfo{person}{S. Liu}, \bibinfo{person}{J. Feng}, \bibinfo{person}{J. Wang}, \bibinfo{person}{Z. Feng}, {and} \bibinfo{person}{et al.}} \bibinfo{year}{2024}\natexlab{}.
\newblock \showarticletitle{Precision HER2: A Comprehensive AI System for Accurate and Consistent Evaluation of HER2 Expression in Invasive Breast Cancer}.
\newblock \bibinfo{journal}{\emph{BMC Cancer}} \bibinfo{volume}{24}, \bibinfo{number}{1} (\bibinfo{year}{2024}), \bibinfo{pages}{1204}.
\newblock


\end{thebibliography}

\end{document}